**Grain-scale thermoelastic stresses and spatiotemporal temperature gradients on airless bodies, implications for rock breakdown.**

Jamie L. Molaro and Shane Byrne, Lunar and Planetary Laboratory, University of Arizona, Tucson, Arizona, USA.

Stephen A. Langer, National Institute of Standards and Technology, Gaithersburg, Maryland, USA.

Corresponding Author: J.L. Molaro, Lunar and Planetary Laboratory, University of Arizona, Tucson, AZ 85719, USA. (jmolaro@gmail.com)

**Abstract**

Thermomechanical processes such as fatigue and shock have been suggested to cause and contribute to rock breakdown on Earth, and on other planetary bodies, particularly airless bodies in the inner solar system. In this study, we modeled grain-scale stresses induced by diurnal temperature variations on simple microstructures made of pyroxene and plagioclase on various solar system bodies. We found that a heterogeneous microstructure on the Moon experiences peak tensile stresses on the order of 100 MPa. The stresses induced are controlled by the coefficient of thermal expansion and Young's modulus of the mineral constituents, and the average stress within the microstructure is determined by relative volume of each mineral. Amplification of stresses occurs at surface-parallel boundaries between adjacent mineral grains and at the tips of pore spaces. We also found that microscopic spatial and temporal surface temperature gradients do not correlate with high stresses, making them inappropriate proxies for investigating microcrack propagation. Although these results provide very strong evidence for the significance of thermomechanical processes on airless bodies, more work is needed to quantify crack propagation and rock breakdown rates.

# 1. Introduction

Thermomechanical processes such as fatigue and shock have been suggested to cause and contribute to rock breakdown in terrestrial environments, particularly in arid environments where other weathering processes are slow [e.g. Eppes et al., 2010; Gomez-Heras et al., 2006; Hall, 1999; Hall and Andre, 2001; Koch and Siegesmund, 2004; Molaro and McKay, 2010; Sumner et al., 2007; Weiss et al., 2003]. The formation and propagation of microfractures in rocks can occur due to expansion and contraction caused by changes in temperature [Todd et al., 1973; Yong and Wang, 1980] and resulting temperature gradients, a process largely controlled by material structure and thermal conductivity. These large-scale temperature gradients can propagate cracks both between and through mineral grains. Cracks can also form and propagate due to mismatches in thermal expansion behavior of adjacent mineral grains [Thirumalai and Demou, 1970; Dey and Wang, 1981]. In this case, the magnitude of these induced stress is proportional to the difference in thermal expansion coefficient between the two grains, multiplied by the Young's modulus and change in temperature, which can lead to tensile stresses on the order of hundreds of MPa [e.g. Evans, 1978; Kranz, 1983; Blendell and Coble, 2006]. These stresses are typically concentrated at grain boundaries [Batzle et al., 1980; Gallagher et al., 1974; Hallbauer, 1973], and can lead to extensional cracks within and between grains. Thermally stressed materials often display grain boundary cracks, which may be continuous along several grains [Kranz, 1979a; Sprunt and Brace, 1974]. As microfractures propagate over time, they may disaggregate near-surface material, increase material porosity, decrease material strength and/or coalesce into larger cracks [Aires-Barros, 1975; Jansen et al., 1993; Kranz, 1979b; Lange, 1968; Luque et al., 2011; Nur and Simmons, 1970; Simmons and Cooper, 1978; Swain and Hagan, 1978; Viles, 2010], breaking down rocks slowly over time.

Airless bodies may provide an environment uniquely suited to this process. For example, slowly rotating bodies that are close to the sun (such as Mercury) experience dramatic temperature ranges and thus large diurnal stresses. Bodies further from the sun (such as asteroids) have much smaller temperature ranges, but rotate quickly and thus experience a more rapid thermal cycling rate. Understanding where in the solar system thermomechanical breakdown may occur has important implications for regolith production rates, surface ages and crater degradation rates, and landscape evolution over time [e.g. Delbo et al., 2014]. It may also provide invaluable information for human exploration endeavors, such as predicting surface properties of future landing sites on different bodies.

Thermal stress breakdown of materials has been suggested as an active process on various bodies in the inner solar system. Jewitt and Li [2010] suggested that the extreme surface temperatures of (3200) Phaethon during its short time at perihelion, only 0.14 AU from the sun, may cause the breakdown of the surface and provide a source of material for the annual Geminid meteor shower. Dombard et al. [2010] suggested thermal processes were responsible for breaking down boulders and creating smooth "ponds" of sediment on the asteroid Eros. They have even been suggested to play a role on smaller bodies, such as small comets [Shestakova and Tambovtseva, 1997; Tambovtseva and Shestakova 1999] and on centimeter- to meter-sized meteoroids [Čapek and Vokrouhlický, 2010]. Viles et al. [2010] found that only modest thermal cycling was needed to cause damage to samples under martian conditions. Other experimental studies have shown that thermal cycling in atmosphere can fracture ordinary, CM, and LL/L chondrites [Levi, 1973; Delbo et al., 2014]. Thermomechanical regolith production rates have yet to be rigorously quantified on these bodies, and how they compare to other processes is yet unknown. Historically, micrometeorite bombardment has been assumed to dominate regolith generation on extra-terrestrial surfaces, with estimated survival times for centimeter to meter-sized rocks on the Moon of $10^6$ -$10^8$ years [Hörz et al., 1975; Hörz and Cintala, 1997; Basilevsky, 2013; Ghent et al., 2014]. Delbo et al. [2014] suggest survival times of material due to thermomechanical processes on near Earth asteroids are comparable to this.

Decades of discussion have explored the role that atmosphere and/or surface moisture plays in crack propagation. Griffith's Criterion [Griffith, 1921] famously states that the theoretical amount of energy required to propagate a crack is proportional to the surface energy of the crack walls at the tip. However, he noticed that the strengths of materials in his experiments were lower in atmosphere than in vacuum. Orowan [1944] attributed this to the presence of fluid within the microcracks. Adsorption of molecules onto the crack walls lowers the surface energy, and thus the amount of energy required to propagate the crack. Later studies have shown that crack velocity is either an exponential or power-law function of the applied stress, and that the velocity is greater by one to several orders of magnitude in the terrestrial atmosphere relative to vacuum (as a function of atmospheric pressure and temperature) [e.g. Atkinson, 1979; Dunning, 1978; Waza et al., 1980]. In a sense, the presence of fluid (either liquid or vapor) affects the localized strength of the material. Typical rock strengths found in laboratory studies in atmosphere are on the order of hundreds of MPa. Estimating strengths of materials on planetary bodies without atmospheres thus presents a challenge.

Thirumalai and Demou [1973] did experiments on lunar analogue samples and found that thermal expansion in a vacuum was the same as in atmosphere, and that both samples displayed permanent, irrecoverable strain caused by grain reorientation from differential expansion. In that sense, vacuum may not limit all thermomechanical activity, but may change its nature on different bodies. Having a suitable way to relate environmental conditions to damage done by thermomechanical processes would be extremely helpful, as we cannot perform laboratory experiments on samples from most parts of the solar system, and many planetary environments are difficult to reproduce.

Many field studies have investigated evidence of thermomechanical breakdown in terrestrial environments [e.g. Eppes and Griffing, 2010; Koch and Siegesmund, 2004; McFadden et al., 2005; Ollier, 1963; Rice, 1976; Siegesmund et al., 2000; Sumner et al., 2007; Viles, 2005; Weiss et al., 2002]. Even in arid environments, however, it is very difficult to separate the contribution of different processes and effects to overall rock breakdown. Other studies have attempted to build models to explain how the process may occur, and what features may result [e.g. Gunzberger and Merrien-Soukatchoff, 2011; Martel, 2006; Molaro and Byrne, 2012; Moores, 2008]. Exploring this thermomechanical behavior of rocks in an environment without the complications of an atmosphere will help us better understand how the process may operate on Earth and other bodies in the solar system.

Much of the more recent work on this topic is based on two particular studies. Todd et al. [1973] conducted laboratory experiments by thermally cycling basalt samples.

They used acoustic emission sensors to measure cracking events in samples subjected to various heating rates, and found almost no events were detected at rates <2 K/min. Richter and Simmons [1974] conducted thermal cycling experiments to measure thermal expansion coefficients of basalt, and concluded that a heating rate of 1-2 K/min should be used "to separate the dual effects of differing thermal expansions of the component minerals *and* the heating rate" on the thermoelastic response of a rock to a change in temperature. They suggested that rapid rates of temperature change could cause crack propagation by setting up macroscopic spatial temperature gradients within the rock, and that crack propagation along grain boundaries occurred due to mismatches in thermal expansion coefficients of the component minerals, but that these two processes occurred in somewhat separate thermal regimes. Thus for them, the threshold served as a filter between two processes. However, over time this rate of temperature change of 2 K/min has unfortunately become an oft-used, overly-generalized threshold for determining whether or not a sample may be experiencing damage from any or from various specific thermomechanical processes in a given environment [e.g. Hall, 1999; McKay, 2009; Molaro and McKay, 2010]. This has produced a body of literature on the topic that is somewhat inconsistent, requiring that we as a community reevaluate our approach.

While misunderstanding and/or misuse of this threshold has propagated through the scientific community over time [Boelhouwers and Jonsson, 2013], recent studies have become less reliant on it. Viles [2010] conducted experiments thermally cycling rock samples under martian conditions. They found that the samples exhibited damage at heating rates much lower than 2 K/min, and concluded that the threshold is an unnecessary requirement for induced strength loss from thermal cycling. The general assumption had been that a temporal temperature gradient sets up a spatial gradient that can cause crack propagation. However, Molaro and Byrne [2012] modeled temperatures of various planetary surfaces at the microscopic scale and found that large temporal surface temperature gradients were not necessarily correlated with large spatial temperature gradients, making it unclear how to reasonably estimate the relative amount of damage expected on each surface. Boelhouwers and Jonsson [2013] instead advocate a reliance on measuring absolute temperature, macroscopic spatial temperature gradients, and actual strain in field and laboratory work. One motivation of the study presented here is to attempt to better understand the relationship between spatiotemporal surface temperature gradients and stress, explore their efficacy as proxies, and try to understand the control they have on thermoelastic behavior of microstructures.

An additional barrier to understanding thermomechanical processes, discussed by Viles [2001], is whether the scale of operation is the same as the scale of observation of that process. If thermal processes are causing breakdown in a given environment, they are most likely to operate (primarily) at the grain scale, since that is where the most extreme effects are induced [Gómes-Heras et al., 2004, 2006]. However, the features we observe may cover a range of scales, not directly revealing the nature of their formation [Cooke and Warren, 1973]. For example, Richter and Simmons [1974] observed permanent deformation in thermally cycled samples by measuring their thermal expansion. This expansion occurs at the rock scale (where it is being measured), but is caused by deformation and reorientation of mineral grains. Similarly, large cracks are observed in boulders, however they form slowly over time due to microcracks propagating between grains, eventually coalescing to form larger scale features [e.g. Jansen et al., 1993]. Field studies that measure temperatures or temperature gradients [e.g. Hall 1999; McKay and Molaro, 2009; Molaro et al., 2010] use thermocouples attached to in-situ samples. In these cases, the thermal forcing is happening at the rock scale, but observation is occurring at the thermocouple scale. With observations, measurements, thermal forcing, and breakdown all happening at different scales, not only does it become difficult to relate measurements that are easy to take (or simulate) in natural environments to observed features, but it also becomes difficult to relate these different studies to each other. This is particularly true for studies of other planetary surfaces, where our observations and measurements are limited by the type and resolution of spacecraft instruments available. It is necessary to link what is occurring at these different scales in order to fully reveal and understand the processes at work, let alone find a suitable proxy for estimating damage produced as a result.

The question that everyone would ultimately like to answer is this: how do we quantifiably relate the external thermal forcing that is easily measured or controlled in the field or laboratory, to the amount and nature of damage rocks experience over time? This study hopes to address a small piece of this very complex puzzle by evaluating thermoelastic stresses that are induced at the grain scale as a result of external, environmental thermal forcing, and explore their relationship with accompanying spatiotemporal surface temperature gradients. Investigating the thermal stress behavior of a microstructure will help us better understand the link between large and small scales, and how measuring temperature gradients might better inform our studies. By modeling stresses in microstructures throughout the inner solar system, we will also gain a better understanding of on which surfaces thermomechanical processes may be important, and the implications that it has for surface evolution over time.

In this study we will present results from modeling grain-scale thermoelastic stresses induced in simple microstructures on various airless body surfaces. We will discuss the effects of mineral grain type and distribution on stresses induced within the microstructure, as well as the effect of pore spaces, and the implications for crack propagation and rock breakdown rates. We will also explore the relationship between grain-scale stresses and spatiotemporal surface temperature gradients, and discuss their usefulness as a proxy for thermomechanical breakdown.

**2. Methods**

In this study, we modeled grain-scale thermoelastic stresses produced on airless surfaces using Finite Element Analysis of Microstructures (OOF2) [Langer et al., 2001]. OOF2 is a 2-D finite-element modeling program developed by the National Institute of Standards and Technology that is designed to help scientists simulate the behavior of microstructures. It has been used extensively in the materials science and engineering communities since its release over a decade ago. In this study, we use OOF2 to model the thermoelastic behavior of microstructures on airless body surfaces, with varying grain sizes and thermophysical properties.

We created mock microstructures for use in this study, as we determined that using micrographs of real planetary materials would be unnecessarily complex at this stage of the research and make the results less informative. Additionally, mock microstructures allow us to independently vary properties such as grain size, composition, and mineral properties. Figure 1 shows two of the microstructures used, each composed of grids of black and white hexagons. The microstructures should be thought of as small portions of an infinite half-space, where the top is the surface that is open to space and the bottom is at depth within the ground. Each hexagon in these microstructures represents a mineral grain, with the black and white colors corresponding to two material types (Table 1). The red diamonds in Figure 1 (b) are pore spaces with no properties assigned, added to approximate the effect of pre-existing micro-cracks. Our treatment of cracks is only approximate, and thus this study does not present a thorough investigation of the effect of cracks within microstructures. It does, however, inform what effect their existence will have on the results presented.

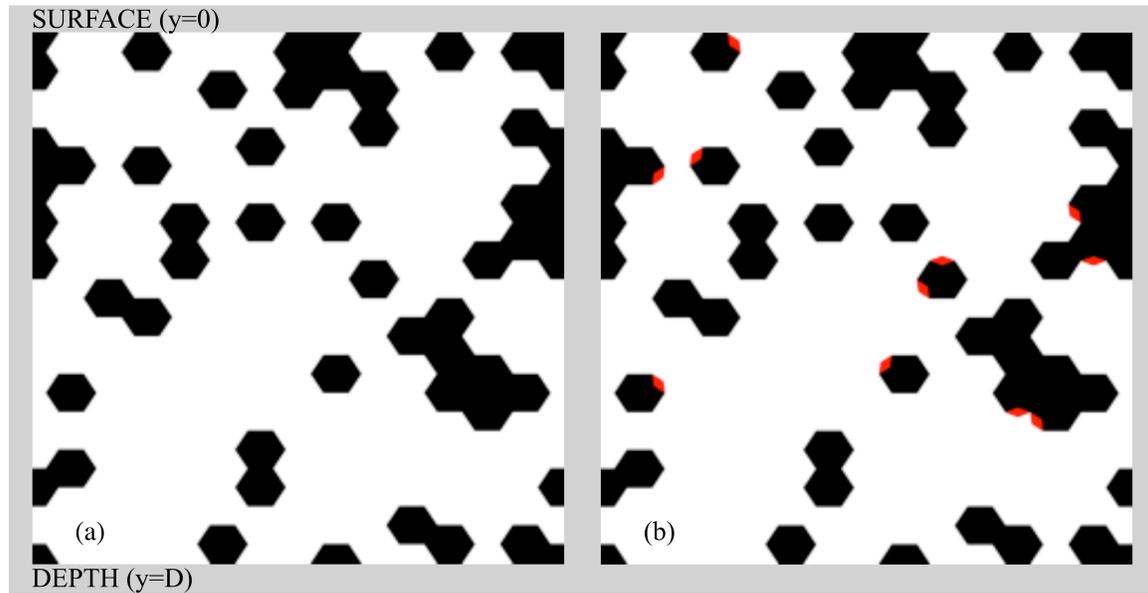

**Figure 1.** Example microstructures composed of black and white hexagons. Black hexagon "grains" are assigned material properties of plagioclase, and white of pyroxene. (a) shows the "standard" microstructure, made of 75% pyroxene and 25% plagioclase. (b) shows the same microstructure with red diamonds, which are assigned no properties.

**Table 1**

| Material Property | Symbol | Pyroxene | Plagioclase |
|---|---|---|---|
| Thermal Conductivity (W/m*K) | $k$ | 4.6 | 2 |
| Heat Capacity*Density (J/kg*K)*(kg/m$^3$) | $\rho c_p$ | 2.7x10$^6$ | 2.9 x10$^6$ |
| Young's Modulus (GPa) | $E$ | 175 | 85 |
| Possion's Ratio | $\nu$ | 0.23 | 0.33 |
| Linear Coefficient of Thermal Expansion (K$^{-1}$) | $\alpha$ | 0.8x10$^{-5}$ | 0.4x10$^{-5}$ |

Typical basalt is made primarily of pyroxene, and 0-50% plagioclase. The microstructures used in this study are made up of some percentage of pyroxene and plagioclase, as listed in Table 2. The microstructure made of 75% pyroxene and 25% plagioclase (Table 2 III, shown in Figure 1) will be referred to as the "standard case" throughout the paper. For each material, we defined values for their density, heat capacity, thermal conductivity, Young's modulus, Poisson's ratio, and coefficient of linear expansion. The literature was searched for relevant material properties for all pyroxene and plagioclase minerals. The material properties for each of the two groups were remarkably consistent, so the average values for each property in each group were taken for this study (Table 1). The full results of the literature search can be found in Appendix Section A1 (Tables A1-A5). While beyond the scope of this study, it is worth noting that some material properties, thermal conductivity in particular, display temperature dependence. The reference temperature for the value of thermal conductivity used is 280 K, close to the mean temperature (288 K) at the lunar surface for this standard microstructure.

In order to evaluate thermoelastic stresses over the course of a solar day on each microstructure, OOF2 must solve the heat (1) and force-balance (2) equations, given by:

$$c_p \rho \frac{dT}{dt} + \nabla Q = 0 \qquad (1)$$

$$M \frac{d^2 u}{dt^2} + \nabla \cdot \sigma = 0 \qquad (2)$$

where $c_p$ is the specific heat capacity, $\rho$ is the density, $T$ is the temperature, $k$ is the thermal conductivity, $Q$ is the heat flux, $M$ is the mass density tensor, $u$ is the displacement field, and $\sigma$ is the stress tensor. The top of the microstructure represents a rock surface, and the bottom is at 5 mm depth ($D$), corresponding to a grain diameter of 360 µm. The boundary conditions applied to the microstructure are as follows:

$$u_x |_{x=0,D} = 0 \qquad (3)$$

$$u_y |_{y=D} = 0 \qquad (4)$$

$$Q_y |_{y=0} = q_o(t) \qquad (5)$$

$$Q_y |_{y=D} = q_D(t) \qquad (6)$$

where $u_x$ and $u_y$ are displacement in the $x$ and $y$ directions, respectively, $D$ is the width and the depth of the microstructure, and $q_o$ and $q_D$ are the time dependent heat fluxes at the surface and bottom of the microstructure, respectively. The assumption here is that the microstructure is embedded in a half-space that is infinite in extent (or at least very much larger than the area being modeled). Thus our model is suitable for extensive exposed rock faces rather than boulder sized objects. Using fixed boundary conditions is appropriate for this model because expansion and contraction is assumed to be isotropic in the horizontal direction, thus holding boundaries within an infinite half-space fixed. To ensure the heterogeneity of the microstructures did not affect this assumption, tests were performed that indicated that rigid and periodic boundary conditions produced comparable results (see Appendix Section A2). This approach has been used for similar studies [Lachenbruch, 1962; Mellon, 1997; Chien et al., 2013] as an appropriate approximation for investigating the thermoelastic behavior of broad surfaces. Mellon et al. (2008) found that this model is appropriate for surfaces on the order of tens of meters for ice, though the analysis has not been completed for rock. This work will thus connect environmental thermal forcing to grain scale effects, but cannot address intermediate scales where the shape, size, and surface curvature of rock may come into play. Note that while the vertical edges of the microstructure cannot move spatially, the microstructures are periodic in the horizontal direction, and heat transfer is periodic across their boundaries.

We used a time dependent heat flux for the horizontal boundary conditions of the model to simulate energy moving into the surface from insolation and thermal emission (5), and conduction out of the bottom of the microstructure into the rock interior (6). These were calculated using a separate 1-D thermal model using equation (1) to calculate the temperature of a much deeper column of material over one solar day on the relevant planetary surfaces. The solar and conductive fluxes entering the surface and exiting the bottom of a layer of this column were stored and used as time dependent boundary conditions in the OOF2 model. The layer thickness and material properties in the thermal model match the depths and bulk material properties for each microstructure in OOF2. Other details on this thermal model can be found in Molaro and Byrne [2012]. It was also used to determine the initial temperatures applied to the microstructure in OOF2. The resulting diurnal temperatures produced by OOF2 in the microstructures are consistent with our thermal model and others within the literature [e.g. Vasavada et al., 1999]. Gómez-Heras [2006, 2008] found that differences in albedos between surface grains do change their relative heat fluxes and thermoelastic relationship to each other, an effect not included in this model. This effect could mean that model results for the surface grains are unreliable, but would likely not affect the remainder of the microstructure. As we will discuss in the results, the highest stresses generated within the microstructure do not occur at the surface, and thus we determined this effect was not important given the scope of this study.

The stress parameter presented in the following results is the equivalent stress ($\sigma_e$) or von Mises stress ($\sqrt{3J_2}$, where $J_2$ is the second invariant of the deviatoric stress tensor). This parameter is calculated from the principal stresses:

$$\sigma_e = \sqrt{\frac{(\sigma_1-\sigma_2)^2+(\sigma_2-\sigma_3)^2+(\sigma_1-\sigma_3)^2}{2}} \tag{7}$$

A material is said to fail when the equivalent stress reaches some critical yield strength of the material. The equivalent stress is commonly used in materials science studies, however it is normally used to evaluate stresses at larger scales (i.e. the scale of a boulder). Typical strengths at these scales are on the order of hundreds of MPa, but are likely higher at the grain scale. In our results we found that the $\sigma_{xx}$ component of the stress is always the principle stress (because the top boundary is a free surface), and that $\sigma_e \approx |\sigma_{xx}|$.

The tensile strengths of materials are much lower than the compressional strengths, and thus tension has the highest potential for contributing to rock breakdown. Indeed, the majority of observed microcracks in laboratory studies [Kranz, 1983] are extensional. Thus for simplicity, we will only focus on the tensile regime, however some of the thermomechanical processes previously described (e.g. mismatches in thermoelastic behavior) may occur in either regime. While the equivalent stress is always positive, a sign correction has been applied to our results (notable in Figures 2, 3, 6, 8,

10) to visually separate tensile (positive $\sigma_{xx}$ values) and compressional (negative $\sigma_{xx}$ values) states. For succinctness, the "equivalent stress" will be referred to simply as "stress" for the remainder of the paper.

## 3. Results

### 3.1 The Moon

The stress state of a microstructure changes dynamically throughout the course of a day. Figure 2 shows the range of stresses induced within a microstructure over one solar day for a flat, equatorial lunar surface. For a homogeneous pyroxene microstructure (black), the range is very narrow and the envelope appears as a thin line. The peak stress in the tensile regime occurs at the surface of the microstructure, just before sunrise. This cannot be seen in the figure, as plotting profiles of surface and bottom stresses would be too close together to distinguish visually. The compressional stress induced in the microstructure increases throughout the entire sunrise and peaks after midday at the microstructure's bottom edge, indicating it is linked more strongly with temperature than spatiotemporal surface temperature gradients. This will be discussed in more detail later.

Adding heterogeneity into the microstructure does not modify this general behavior, but does change the amplitude and distribution of induced stresses. The peak tensile stresses still occur pre-sunrise, however not necessarily at the microstructure surface. Figure 2 (green) shows the range of stresses induced in a standard microstructure throughout the solar day, which is much larger than the homogeneous case. The average stress experienced throughout the microstructure decreases (Figure 3, green dashed) with added heterogeneity, however the maximum stress induced increases, peaking at 150 MPa (Figure 3, green solid). A snapshot of the microstructure during the state of peak tension (Figure 5 III) reveals that the highest stresses are, indeed, not induced at the surface as in the homogeneous case, but are scattered throughout. This indicates it is not just the temperature, but also the heterogeneity of the material that controls thermoelastic behavior.

Figure 4 shows the average and maximum stresses induced within the microstructure during the state of peak tension. The values for each data point labeled on Figure 4 are listed in Table 2 for reference. The average stresses (black line) show a linear decrease moving from a homogeneous pyroxene to a homogeneous plagioclase microstructure, where the slope is determined by the difference in stress between the two end-members. The stresses in the homogeneous cases are controlled by the individual material properties of each mineral. The peak stress of 32 MPa in (VII) contrasts strongly with the 97 MPa in (I). Compared to pyroxene, plagioclase has a lower Young's modulus and coefficient of thermal expansion, and thus experiences less stress as a function of changing temperature. The contribution of different material properties to thermoelastic behavior is discussed more below.

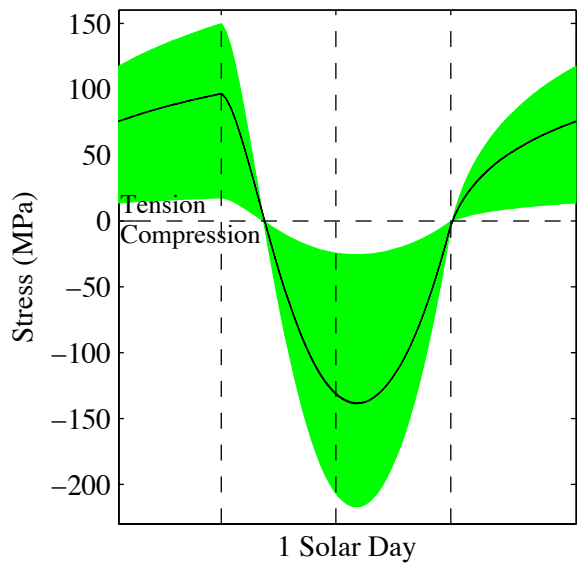

**Figure 2.** Profile of the range of stresses within a microstructure over one solar day for a flat, equatorial Lunar surface. The black envelope represents a homogeneous pyroxene microstructure and the green, a microstructure with 25% plagioclase and 75% pyroxene grains. The vertical dotted lines represent the time at which sunrise, noon, and sunset occur, from left to right.

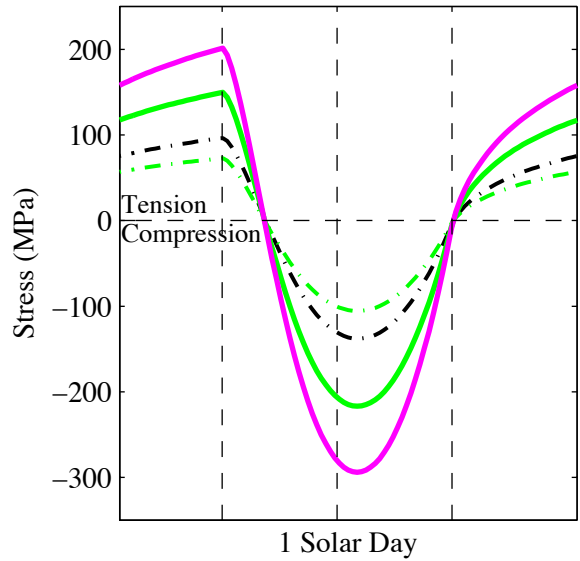

**Figure 3.** Profiles of stress over one solar day in lunar microstructures. The dashed lines are the average stresses in a homogeneous pyroxene (black) and a standard (green) microstructure. The solid lines are peak stresses in a standard (green) microstructure, and a standard microstructure including cracks (red).

Unlike the average stresses, however, the maximum stresses (Figure 4, green line) jump immediately when only a small amount of heterogeneity is added. For example, the jump of 36 MPa from the homogeneous case (I) to (II) with the addition of only a single plagioclase grain shows that the difference in elastic properties between mineral grains dominates the amplitude of induced stresses in heterogeneous microstructures. Case (VI) also displays this jump from (VII) very dramatically, with an increase in stress of 52 MPa. The maximum stress in (VI) is also close to the value in (I). This suggests that any heterogeneous microstructure will experience stress at least approximately equal to the maximum stress induced in a homogeneous microstructure of whichever mineral component has the highest thermal expansion coefficient and Young's modulus (in this case, pyroxene). Results of additional tests conducted to support this suggestion can be found in Section 3.3 and Appendix Section A3.

Figure 5 compares snapshots of the peak tensile stress state for microstructures (II) through (IV) from Table 2. Panel (II) shows that high stresses are concentrated along surface-parallel boundaries of the grain. This is expected, as the differential expansion and contraction of minerals in the x-direction "pulls" along those boundaries where adjacent grains meet. An increase in the heterogeneity in (III) yields a moderate increase to the maximum stress. This occurs due to amplification of stresses in areas where fields from clustered surface-parallel grain boundaries interact, setting up complex stress fields within the pyroxene. One might expect the stress at a plagioclase grain where no amplification occurs from nearby grains (e.g. the grain indicated in panel 5 III) to be comparable in magnitude to that of the individual grain in case (II). However, these more isolated grains typically have stresses of ~106 MPa, which is lower than in case (II), and only slightly higher than the homogeneous stresses in (I). This indicates that stress fields can also interact to reduce stresses. Overall, this suggests that the path a cracks takes through a microstructure is controlled by local amplification due to its unique grain distribution. This may have important implications for how rocks disaggregate, which will be addressed in the discussion section.

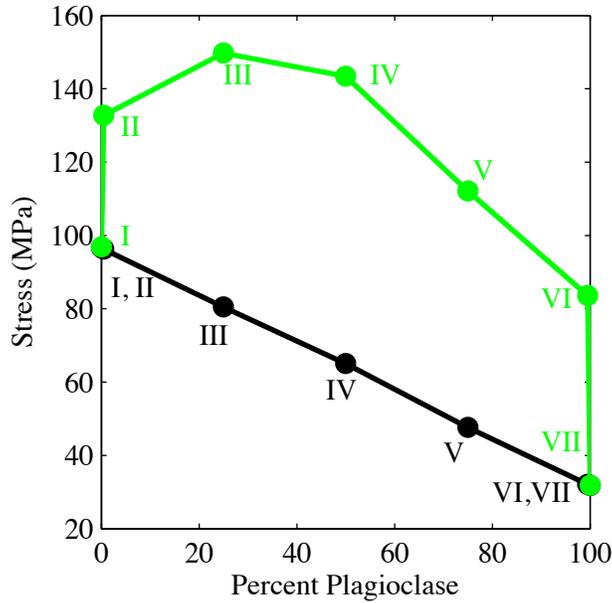

**Figure 4.** The average (black) and maximum (green) stresses induced during the state of peak tension in a microstructure with a given percentage of pyroxene and plagioclase grains. The values for each point on the plot are given in Table 2.

Table 2

| Microstructure | % Pyroxene | % Plagioclase | Peak Average (MPa) | Peak Maximum (MPa) |
|---|---|---|---|---|
| I | 100 | 0 | 97 | 97 |
| II | 99.995 | 0.005 | 96 | 133 |
| III | 75 | 25 | 73 | 150 |
| IV | 50 | 50 | 65 | 143 |
| V | 25 | 75 | 47 | 112 |
| VI | 0.005 | 99.995 | 32 | 84 |
| VII | 0 | 100 | 32 | 32 |

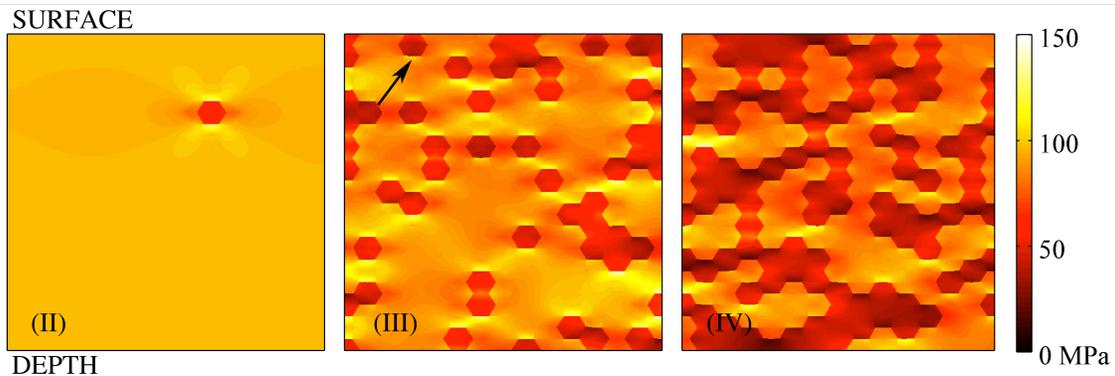

**Figure 5.** Snapshots of the state of peak tensile stress in microstructures corresponding to rows II-IV in Table 2. Center panel (III in Table 2) is the standard microstructure. The microstructures have a domain of 5 mm and grain diameter of 360 μm.

The further increase in heterogeneity in (IV) and (V) shows a decreased peak stress. This suggests there may be an optimal relative volume of mineral grains within a microstructure for producing the highest stresses, which is likely dependent on the mineral types that make it up. This effect is not a consequence of the randomized grain distribution, as shown in Appendix Section A4. However, it is unknown to what extent this effect may be present with a more realistic microstructure that includes complexities such as defects, pore space, and additional mineral types.

The addition of "pre-existing damage" into these microstructures shows an increase the stresses induced. Figure 6 (a) shows a snapshot of the homogeneous pyroxene microstructure with cracks during the time of peak tensile stress. Stresses are concentrated at the tips of the cracks, and are highest at crack tips perpendicular to the surface. The peak stress in this microstructure is 222 MPa. This is a significant increase to that of a microstructure without cracks (150 MPa), however it is much lower than what is predicted by theory. This is likely because diamond-shaped finite elements are an imperfect representation of actual microcracks, which are long and thin. According to Griffith's theory of crack propagation, the stress induced at the tip of an elliptical microcrack is proportional to two times its aspect ratio (2 x length/width). The aspect ratio of the red diamonds is only ~5, where that of a real microcrack is $10^3$-$10^5$ [Kranz, 1983]. This does not necessarily mean that the stresses seen in this study should be expected to increase by an order of magnitude or more, only that they represent a lower limit on realistically expected values, and demonstrate that an increase due to the presence of the crack will occur. These model runs highlight the expected general thermoelastic behavior of a microstructure that contains microcracks.

Note that while one crack does appear to intersect the microstructure surface, we determined that the behavior it produced was unphysical. The elements in the mesh at the tip are still connected, even though in a realistic situation a crack would permanently sever that connection. Occasionally other anomalies seemed to occur, most likely due to imperfect mesh elements when setting up the model run. All of these were excluded from the data during analysis.

Figure 3 shows a profile of maximum stress for the standard microstructure without (green solid line) and with (magenta solid line) cracks, the latter showing a peak stress under tension of 202 MPa. Panel 6 (b) shows the standard microstructure at the time of peak tension with small cracks included. Stresses are still concentrated along surface-parallel grain boundaries, but also additionally at crack tips, particularly where they intersect those boundaries. Stresses at crack tips in a homogeneous case (Panel 6, a) reveal higher stresses than for crack tips in the standard case (Panel 6, b). This is likely because the plagioclase grains in the standard case strain more easily and can take up some of the stress that would otherwise be experienced.

Figure 6 (c) and (d) show longer cracks in a homogeneous and standard microstructure, respectively. The cracks in these cases are two and three times longer than those in A and B. As seen with stress fields arising from heterogeneity, the fields of cracks can intersect in complex ways as well. For example, the two cracks in the lower left (1) induce an amplified field between them. In a real microstructure, it is likely that cracks in that situation would eventually coalesce. The crack in the upper right (2) has a higher stress than the others, likely an artifact due to the fact that its stress field is strongly interacting with a boundary that is rigid. Comparing the two the cracks at (3) and (4), we observe that (4) has a higher stress due to the fact that it is fully perpendicular to the surface. As with the shorter cracks, those perpendicular to the surface induce higher stresses than those parallel. Comparing crack (5) to (4), we should see that it has a lower stress because it is shorter, and thus has a smaller aspect ratio. However, these results show stresses comparable in value at each crack tip. This is most likely because while the cracks are longer, the shape of the tip itself did not change. Due to practical considerations, it is difficult (and computationally expensive) to generate element meshes with very thin elements. Future work will look at this effect in more detail.

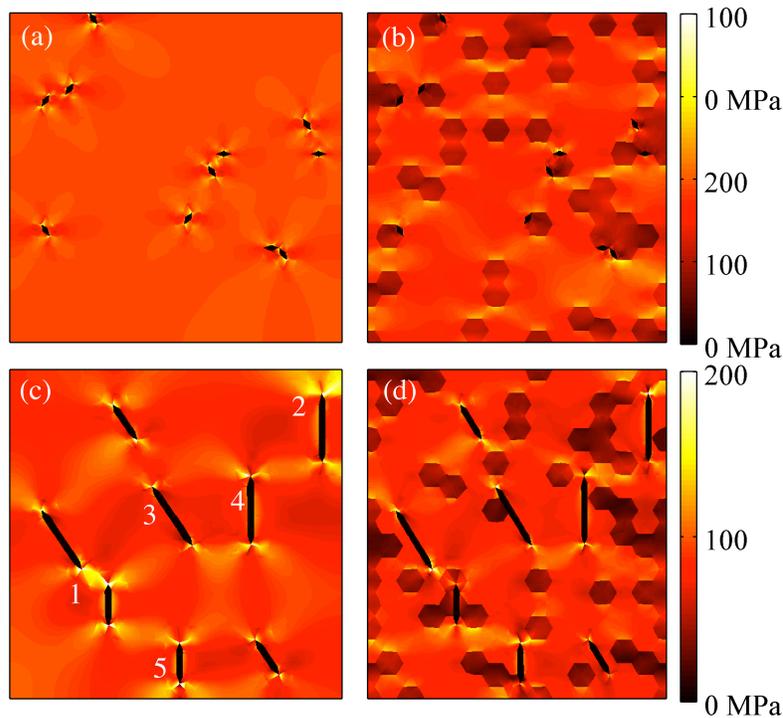

**Figure 6.** Snapshots of the state of peak tensile stress in four microstructures. (a) and (c) are homogeneous microstructures with short cracks, and cracks of 2 and 3 times in length, respectively. (b) and (d) are the same in standard microstructures. The peak stresses at the crack tips labeled in (c) are (1) 295, (2) 292, (3) 207, (4) 226, (5) 233 (at top tip) MPa. The microstructures have a domain of 5 mm and grain diameter of 360 μm.

Flat surfaces on airless terrestrial bodies are typically covered in at least a thin layer of dust or regolith. Regolith has a low thermal inertia, and as a result, it tends to thermally insulate the subsurface. The results of this model approximate the behavior expected for bedrock with no regolith cover, which realistically on the Moon would be a more highly sloped surface. Profiles of maximum stresses induced in highly sloped (65°) east- and west-facing standard microstructure (blue lines) compared to the flat case (green line) are shown in Figure 7. East- and west-facing surfaces have peak stresses of 133 MPa and 127 MPa, respectively, or a decrease from the standard case of 17 MPa and 23 MPa, respectively. This is a result of the fact that sloped surfaces see less open sky than flat terrain, and thus do not cool as efficiently. While these are non-negligible differences, given the uncertainty in actual yield strengths of materials, it is unclear to what extent they may be relevant. The sloped surfaces experience more stress when the microstructures are under compression, however the compressional strengths of materials are higher than tensional strengths, and thus even the difference of 66 MPa seen in Figure 7 may not be important in this context.

Studies can be found in the literature (e.g. Richter and Simmons 1974) that indicate that rocks with a larger grain size are weaker. No appreciable differences in stresses induced in microstructures with varying grain sizes were detected in the limited model runs done to explore this effect. Future work will look at this effect in more detail.

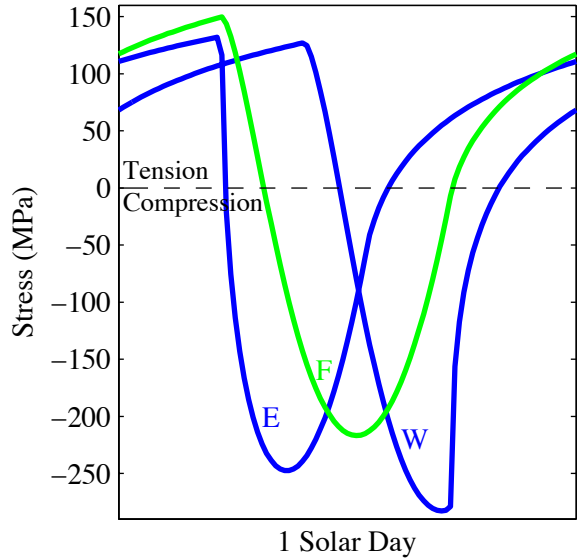

**Figure 7.** Profiles of maximum effective stresses induced in east- and west-facing standard microstructures with a slope of 65° (blue) and of 0° (green line).

**3.2 Elsewhere in the Inner Solar System**
Stresses on lunar surfaces are comparable to typical yield strengths of rocky materials, suggesting that thermal stress weathering may be effective on the Moon. However this may not necessarily be true for other bodies. With a much shorter solar day length (~5.5 hours) and much longer solar distance, bedrock surfaces on Vesta are very cold (~200 K) and temperatures vary little throughout the diurnal cycle. Figure 8 shows the profile of maximum stress in an undamaged standard microstructure on Vesta (magenta line), yielding a peak of 5 MPa. Phobos is closer to the sun, but only shows a maximum tensile stress of 13 MPa. Phobos is a somewhat unique case because it experiences a solar eclipse of 54 minutes at midday each rotation near martian equinoxes (no eclipses occur near solstices). This induces a second spike of tensile stresses during martian equinox seasons that peaks at 6 MPa. Both Phobos and Vesta experience very rapid thermal cycling rates, with Phobos effectively having two cycles per solar day (~7.7 hours) during much of the martian year. However, while these are lower limits on the stresses that could be produced on these surfaces, rock breakdown is less likely to be effective on Vesta and Phobos than on the Moon.

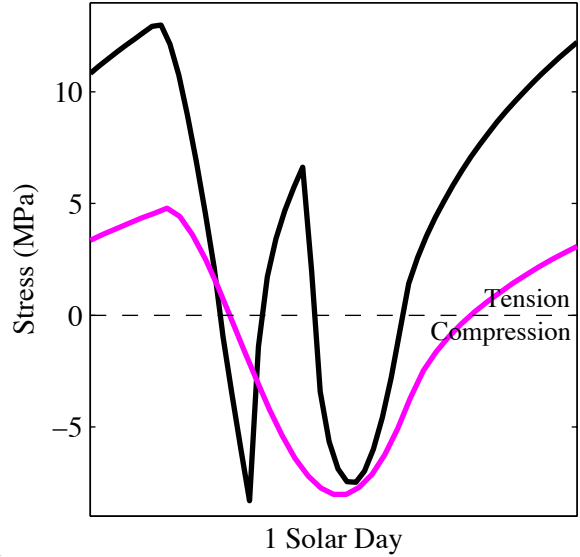

**Figure 8.** Maximum effective stress over one solar day for a microstructure on Vesta (magenta) and Phobos (black). The two profiles are normalized to 1 solar day, even though Vesta has a solar day length of ~5.5 hours, and Phobos ~7.7 hrs.

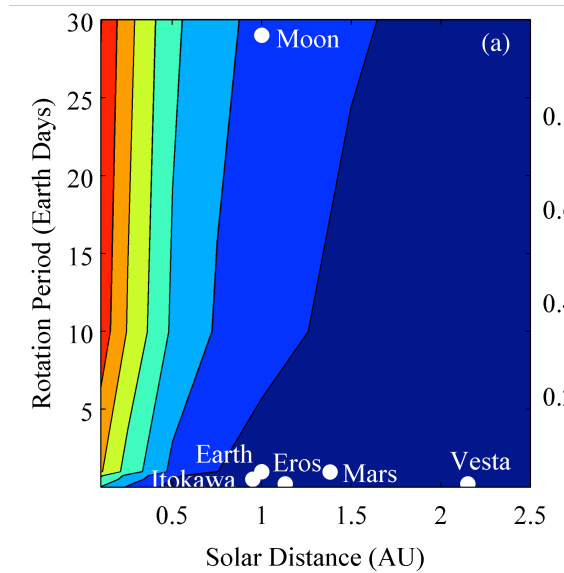

**Figure 9.** Maximum effective stress during the state of peak tension on arbitrary surfaces of bodies with varying solar distance and solar day length in the inner solar system (A), with contour lines spaced at 100 MPa intervals. (B) shows a closeup of rapidly rotating bodies, with contour lines spaced at 50 MPa intervals. All bodies used for these model runs were assumed to have zero eccentricity and inclination.

Other bodies in the solar system, particularly asteroids, provide a variety of orbital properties that likely represent a range of thermoelastic stress regimes. Figure 9 shows peak tensile stress on arbitrary bodies with varying solar distance and solar day length. The standard microstructure was used in each case. The bodies were assumed to have zero obliquity and eccentricity. The general trend shown in Figure 9 (a) shows that bodies that are close to the sun and bodies with long rotation periods induce the highest thermoelastic stresses, and thus are most likely to be susceptible to breakdown from these processes. The perihelion positions of notable solar system bodies are marked on the plots. Some asteroids, such as Phaethon and Icarus, have very close approaches to the sun, and thus are expected to experience very high stresses (in addition to their rapid cycling rate). Most notable among slowly rotating bodies are the Moon and Mercury, however there is a known group of slow rotating asteroids as well [e.g. Pravec et al., 2008]. All of these bodies are good candidates for active thermal breakdown processes.

**3.3 Temperature Gradients and Material Properties**

Spatial and temporal surface temperature gradients ($\nabla T$ and $dT/dt$, respectively) are sometimes used as qualitative proxies for stress. Part of the goal of this study is to explore how these temperature gradients relate to the actual magnitude of stresses induced in a microstructure. In this work, we do not consider additional stresses imposed by any macroscopic (on the order of a thermal skin depth) thermal gradients that might be present. Here we investigate the correlation of thermal gradients at the individual grain scale and across collections of several grains (domain-scale) with stress. Figure 10 (a) shows the average stress in a standard microstructure with temperature gradient for one solar day on the Moon. The temperature gradient is calculated from the difference in average temperature at the top (surface) and bottom (5 mm depth) of the microstructure. The highest magnitude $\nabla T$ occurs during midmorning and sunset, whereas the highest magnitude stresses occur at noon and just before sunrise. Given the geometry of the model, a negative gradient occurs during cooling. This would lead us perhaps to expect high tensile stresses to be correlated with large negative gradients, however this figure indicates that is not the case. Figure 10 (b) shows the same plot for $dT/dt$ (calculated from the thermal model) and average stress, revealing a similar pattern. The times of day at which rapid changes in temperature occur are not the times of day in which the highest stresses occur. Panel (c) shows the same plot, but for a microstructure on an east- (cyan) and west-facing (magenta) slope. In these curves, a spike in $dT/dt$ is easily observed when sunrise (east) and sunset (west) occur, however that spike is not associated with a corresponding spike in stress. The lack of correlation between high $\nabla T$ and $dT/dt$ and high stresses suggests that their use as a proxy is misleading when discussing grain-scale processes.

A snapshot of $\nabla T$ at peak tension (Figure 11, a) shows that the largest $\nabla T$ values are concentrated primarily within the plagioclase grains. This is due to plagioclase's low thermal conductivity. Since they have lower thermal conductivity, strong temperature gradients are setup across the plagioclase grains, while the heat becomes uniformly distributed within the pyroxene more quickly. However, these plagioclase grains have lower stresses than other parts of the microstructure. Figure 11 (b) shows that stress and $\nabla T$ are, in fact, anti-correlated. Figure 12 displays the relative importance of different material properties to induced stresses. In each case, all grains were assumed to have the same material properties (an average of the two values in Table 1), except one property that remained different. The first column in Figure 12 has grains with different thermal conductivity. Columns two and three showcase Young's modulus and Poisson's ratio, and coefficient of thermal expansion, respectively. Snapshots of stress (Row 1) and $\nabla T$ (Row 2) are shown for each case. These figures show very dramatically that while thermal conductivity dominates $\nabla T$, it has virtually no control over induced stress. The opposite is true for Young's modulus/Poisson's ratio and coefficient of thermal expansion. The former appears to have the strongest control on the difference in stress between mineral types, and the latter on generating the highest overall magnitude stresses.

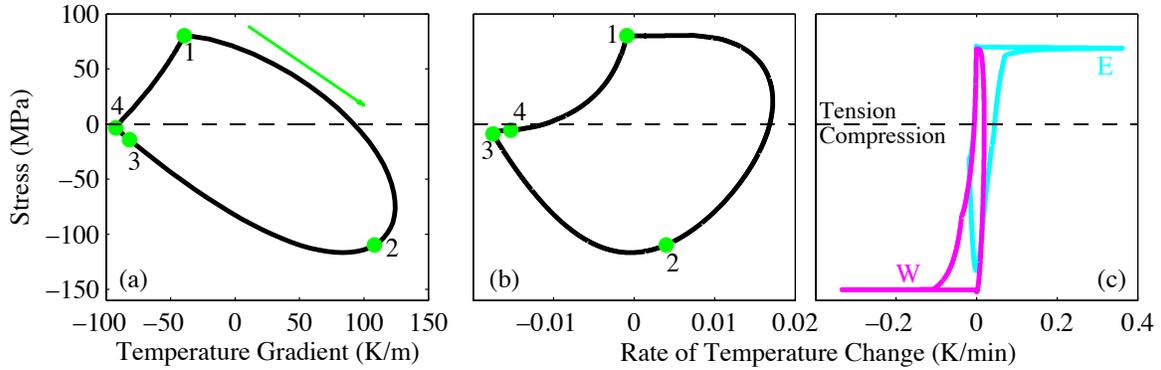

**Figure 10.** (a) Average stress in a standard microstructure with vertical temperature gradient over one solar day. (b) Average stress in a standard microstructure with rate of temperature change over one solar day for a flat surface. (c) Same profile as in (b) for an east- (cyan) and west-facing surface (magenta) with a slope of 65°. The panels (a) and (b), point 1 is the beginning of sunrise, 2 is noon, and 3 and 4 are the beginning and end of sunset, respectively.

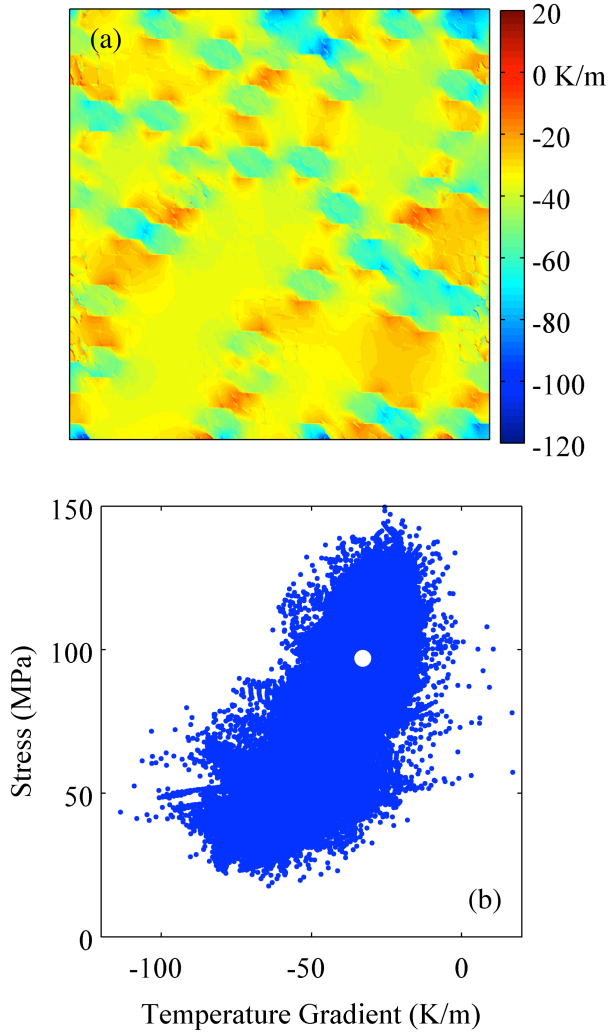

**Figure 11.** (a) snapshot of the temperature gradient within a standard microstructure at the time of peak tension. (b) scatter plot of temperature gradient with stress. The white circle is the value if the average stress and temperature gradient in a homogeneous microstructure, as from Figure 10 (a). The microstructure has a domain of 5 mm and grain diameter of 360 μm.

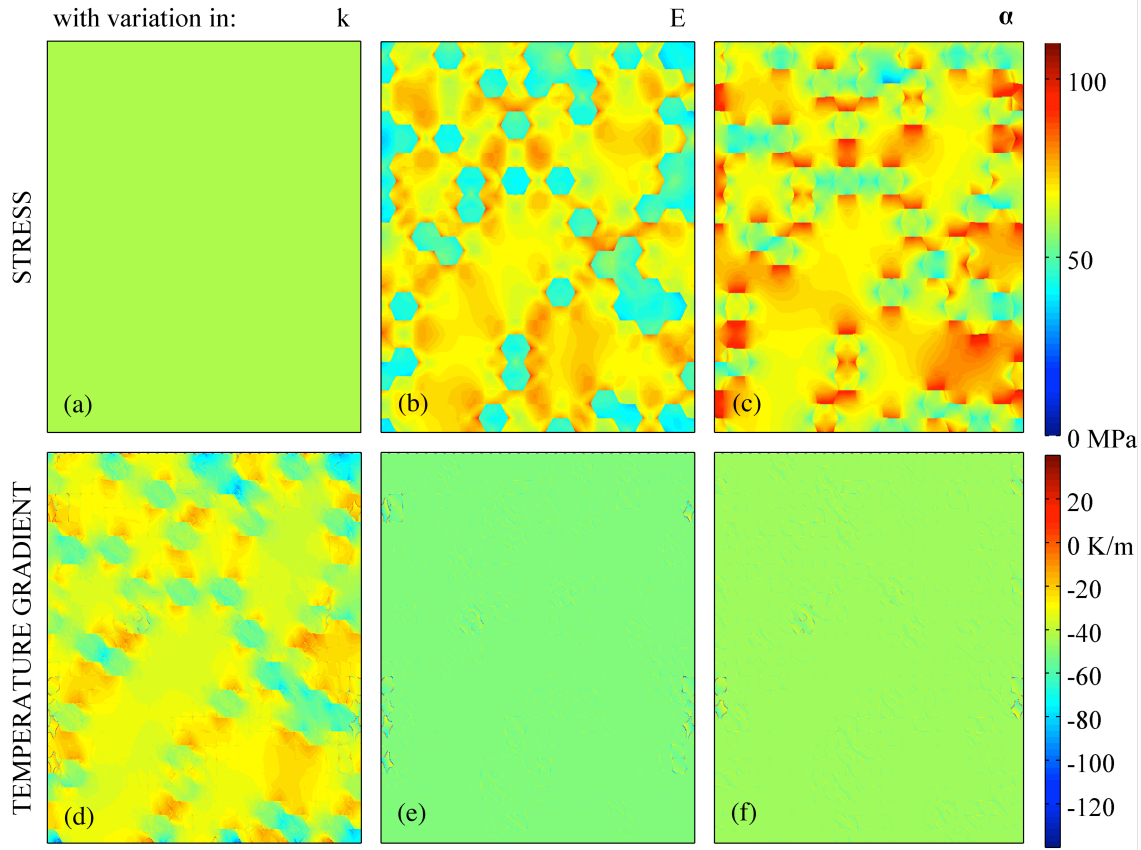

**Figure 12.** Snapshot of the stress (a-c) and $\nabla T$ (d-f) during the time of peak tension in a standard microstructure. The mean of the Pyroxene and Plagioclase values of all material parameters listed in Table 1 are assigned to all grains, with the exception of a single parameter remaining different in each test. The thermal conductivity values remain different between the mineral grains in the first test (left column, panels a and d), Young's modulus and Poisson's ratio in the second test (middle column, panels b and e), and coefficient of thermal expansion in the third test (right column, panels c and f). The peak values for each panel are a) 60 MPa, b) 95 MPa, c) 106 MPa, d) -143 K/m, e) -46 K/m, and f) -46 K/m. The microstructures have a domain of 5 mm and grain diameter of 360 μm.

## 4. Discussion

### 4.1 Stress Distributions and Amplitudes

The findings reported here are consistent with thermoelastic behavior of rocks found in the literature. According to theory, stresses between mineral grains are on the order of the difference in thermal expansion coefficient between the two grains, multiplied by the Young's modulus and change in temperature [Kranz, 1983]. The results presented agree with this rule of thumb, but being able to view their distribution within the microstructure provides additional insight into how thermomechanical processes may operate, and what factors ultimately control rock breakdown. It is important to keep in mind that these stresses represent idealized amounts of energy available for crack propagation, and they do not account for the release of mechanical energy when propagation does occur. There are also natural effects that may modify their magnitude, some of which are discussed below. The distribution of stresses presented is also consistent with observed behavior of samples in laboratory studies, as discussed in the following sections.

Our results indicate that temperature and composition control the amplitude of induced stresses in microstructures on planetary surfaces, with the coefficient of thermal expansion as the most dominant material property. This is unsurprising, as the volume change during thermal expansion or contraction is directly proportional to the change in temperature. Young's modulus also plays a significant role, however the thermal conductivity contributes very little. A homogeneous microstructure will experience some baseline stress determined by these properties. With the addition of heterogeneity, grains of both materials still experience intragrain stresses close to their mineral type baseline, even when mixed. Therefore the average stress over the microstructure is determined by the percentage of each mineral grain present, and what their baseline values are. In other words, adding pyroxene to a homogeneous plagioclase microstructure will raise the average stress because pyroxene has a higher baseline value. The implication of this is that even by adding a single pyroxene grain, some part of the microstructure will experience a stress comparable to pyroxene's baseline value. The minimum stress within the microstructure during this state of tension is thus determined by which mineral has a higher baseline stress value. Thus, the peak tensile stress experienced by a microstructure with any heterogeneity is at least comparable to the baseline stress value of the material with the highest coefficient of thermal expansion. This can be used as a method of approximating stresses in situations where modeling the full solution is difficult, such as on Mercury where the insolation cycle is very long. Assuming its microstructures contain pyroxene, this method yields a lower limit for the peak tensile stress on Mercury of 350 MPa.

The existence of heterogeneity (even in tiny amounts) provides the strongest control on increases over the baseline stresses of each mineral. Additional increases also occur due to concentration of stress at surface-parallel grain boundaries as a result of the difference in thermoelastic behavior between adjacent grains. In areas where stress fields from these grain boundaries are clustered, those fields interact and often are amplified. Stress amplification does vary somewhat with relative percentage of mineral types, however it is not a strong effect. The fact that the stresses interact strongly where grain boundaries are clustered indicates that the thermoelastic behavior of individual rocks will be unique, and highly dependent on mineral types and distributions. Including pre-existing damage or pore space to these microstructures also increases maximum stresses, though these results are very limited in exploring to what extent. While we have demonstrated that an increase does occur, it is unclear what increase might be expected with a more realistic microcrack size and shape.

In the homogenous cases, surface-perpendicular cracks induce higher stresses than surface-parallel cracks. This is reasonable because most of the stress is in the x direction and can take advantage of narrow crack tips in a surface-perpendicular crack. However, the heterogeneous cases do not display this orientation preference as strongly. Instead, the highest stresses are induced where crack tips intersect concentrations of stress along adjacent mineral grain boundaries, especially in areas where these boundaries are abundant, or when fields from crack tips interact with each other.

It is worth noting that the compositions of many planetary surfaces are not well characterized. On the Moon, pyroxene and plagioclase are common, and vary in relative abundance between the mare and the highlands. Volcanic rocks on other terrestrial planet surfaces will contain some amount of these minerals as well. However, the lithology of asteroidal rocks may be dominated by other minerals. For example, ordinary chondrites are primarily composed of pyroxene, olivine, and iron [McSween et al., 1991]. Carbonaceous chondrites also have a wide variety of minerals, including carbonates, sulfates, phosphates, and iron/nickel compounds [Zolensky et al., 1993]. These mineralogies will determine unique stress amplitudes and distributions for individual rocks. Nevertheless, the baseline stress values and/or methodology presented in this study can still be used to approximate lower limit stresses. In reality, stresses are likely much higher since the presence of defects and microcracks in a real microstructure can cause high stress concentrations.

**4.2 Implications for rock breakdown**
These results have interesting (although not straightforward) implications for how cracks propagate through materials and how rocks actually breakdown and disaggregate. Cracks will tend to propagate along the "local maximum stress trajectory" [Kranz, 1983; Wu et al., 1978] in order to maximize the strain energy released. Since stresses at crack tips are highest for surface-perpendicular cracks, this would suggest it is energetically favorable to propagate vertically toward the microstructure surface. Areas of high stress amplification due to heterogeneity may serve as regions that draw cracks, causing them to twist and turn as they propagate upward. If these areas are close to each other, they may also allow for the coalescence of microcracks, releasing more strain energy. The grain distribution will likely strongly influence the resulting size and shape of disaggregated material, since the local stress amplifications could set up naturally preferred crack spacing.

Some effects can arrest propagation, such as interacting directly with a grain [Kranz, 1983]. A vertically propagating crack could terminate at the bottom of a grain as it moves towards the surface. On the other hand, since stresses concentrate along surface-parallel intergrain boundaries, a crack intersecting the bottom of a grain may curve around its edge and continue upward. If the grain was at the surface, this could result in granular disintegration. The presences of pores and microcracks will also modify and/or halt propagation by altering local stress fields or local cohesive strength [Kranz, 1983; Walsh and Lomov, 2013]. Extensional cracks within grains can form to relieve stress, and in some cases even transgranular cracks [Kranz 1983]. Other effects not included in this model, such as from surface curvature [Martel, 2006], will also affect crack propagation and material breakdown.

**4.3 Material Strength and Porosity**
The ability for crack propagation to occur ultimately depends on material strength, which is variable and depends on attributes such as material composition, size, shape, porosity and/or microcrack density, and confining pressure [e.g. Cotterell et al., 1995; de Castro Lima, 2004; Hasselman, 1969]. In discussions of material strength, researchers tend to think about the strength of a rock as a whole. This is in large part due to the fact that many of the studies on this topic are done for engineering purposes where large scales are more important. Strengths are determined by how much pressure it takes to crush (compressional) or pull apart (tensional) a sample. Typical rock strengths in this context are on the order of tens to thousands of MPa, and tend to be lower in larger grained and/or granitic materials [Jaeger and Hoskins, 1966; Mellor and Hawkes 1971; Nur and Simmons, 1970; Simmons and Cooper, 1978; Ersoy and Atici, 2004]. Measurements have also been made of the fatigue strength of materials, typically on the order of hundreds of MPa [e.g. Krokosky and Husak, 1968]. However, strength and failure in this context are not well defined, as fatigue is a macroscopic measurement that encompasses a variety of microscale processes. Unfortunately, none of these measurements are suitable for determining the stress required to propagate an individual crack.

Beyond the question of what the material strength is, there is the question of what happens over time as a rock is thermally processed to some extent. The thermal shock resistance of rock samples has been shown to decrease with increasing porosity [Coble and Kingery, 1955; Sousa, 2005]. However, while the strength is decreasing, accumulated microcracks also reduce the effective Young's modulus of a material, thus the amount of stress experienced during a change in temperature [Cotterell et al., 1995; Hasselman, 1969]. This would suggest that damage accumulation is non-linear and slows over time. Thermal fatigue has also been shown to reduce the overall strength of rock samples over time [e.g. Jansen, 1993; Mahmutoglu 1998], however as a process it operates at multiple scales. On one hand, movement of microcrack walls can absorb energy that would otherwise be available to propagate cracks [Atkinson, 1984]. However, if the stress concentrated at a crack tip is high enough to overcome local material strength, then the growth of the crack will release mechanical energy, and additional stress can be relieved by minor reorientation of grains into these pore spaces that are created over time. Early studies of crack propagation indicate that the coalescence of microcracks does not occur if the cracks are more than a few crack lengths apart [Bombolakis, 1964, 1968, 1973; Hoaek and Bieniawski, 1965], however if locally extensive enough, coalescence can drive development of macroscopic cracks [David, et al, 1999; Jansen 1993] within planes of weakness that would be able to take advantage of an effective Young's modulus lowered by the increase in porosity. In this sense, it is difficult to tell how microcrack growth rates vary over time.

The results reported here show an increase in stress at the tips of pore spaces, which serve as sites where microcrack propagation and lengthening could occur. Theoretically, the stress concentrated at the tip of a crack can be several orders of magnitude higher than the stress away from the crack. However, in light of the discussion above, it is difficult to determine how much of this idealized stress will actually be available to propagate cracks, or what material strength is reasonable to expect. We can use Griffith's criterion [Griffith, 1921] to determine the critical stress ($\sigma_c$) at which a crack will propagate. The criterion states that the amount of energy it takes to propagate a crack is the amount of surface energy the additional new crack walls will have ($\sigma_c=(2E\gamma/\pi)^{1/2}$ where $E$ is Young's modulus, and $\gamma$ is the crack wall surface energy). If we use the Young's modulus of pyroxene (175 GPa), and a surface energy on the order of 1 J/m$^2$ (as used in his original experiments), this yields a critical stress of ~10 GPa. As discussed in the introduction, the surface energy is substantially higher in vacuum than in atmosphere, and will also be affected by crack length and local composition and structure in a variety of ways. Further research is needed to better estimate strength and failure expectations of planetary materials. In particular, laboratory studies of planetary materials in vacuum would be informative and helpful for comparison with models.

Models and macroscopic observations indicate that existing microcracks subjected to a hydrostatic confining pressure of 100 to 200 MPa decreases crack porosity [Cheng and Toksoz, 1979; Feves and Simmons, 1976; Hadley, 1976; Siegfried and Simmons, 1978; Wang and Simmons, 1978; Warren, 1977], however extensive observations of partially healed or filled cracks indicate that microcracks are not completely annealed and accumulate debris over time that prevents them from closing completely [Batzle and Simmons, 1976; Cox and Etheridge, 1983; Padovani et al., 1982; Richter and Simmons, 1977; Shirley et al., 1978; Sprunt and Nur, 1979; Wang and Simmons 1978]. This suggests that microstructures in large rocks or outcrops (as opposed to small rocks that can expand and contract freely), as we model here, may experience partial healing during the day when subjected to large compressional stresses. Those bodies that are subject to the highest potential tensile stresses are thus also subject to the highest compressional stresses (Figure 9). This could decrease breakdown rates on some bodies, in a sense increasing effective strength. Laboratory studies on larger samples would also be informative, but necessarily are more difficult and expensive to do.

## 4.4 Time and Space

These model runs approximate what is going on inside a small element of a much larger system. The stresses induced in microstructures throughout a rock attenuate with depth. The magnitude of the stress values depends on the material parameters, but the skin depth (given by $\delta=(kP/\pi\rho c)^{1/2}$ where $P$ is the diurnal period) for peak stress is the same as for peak temperatures [Holzhausen, 1989]. For example, if the maximum stress during the time of peak tension with depth in a lunar material is ~150 MPa, the stress at 5 skin depths is only a few MPa. If we assume that a crack will propagate at a stress of, say, 100 MPa, then the penetration of damaging stresses is ~0.5 m on the Moon. The cumulative amount of time the near-surface spends in a damaging state is ~33% of the lunar day. This time spent under these conditions will have a significant affect on long-term breakdown rates. This example, however crude, simply demonstrates that thermomechanical processes likely only effect the near-subsurface, and will be dependent on day length. Other stresses caused by edge and shape effects of rocks could cause thermomechanically-activated processes at greater depths, however those effects are beyond the scope of this model.

## 4.5 Remote Sensing and Temperature Gradients

While spatial temperature gradients at the macroscopic scale may be useful [Čapek and Vokroulichý, 2010] for investigating macroscopic processes, our results suggest that at grain scale they are not. Neither the grain- or domain-scale temperature gradient correlates with the location or time of peak thermal stresses within a microstructure. This indicates that grain or multi-grain scale spatial temperature gradients are not an appropriate proxy for evaluating microscopic thermoelastic stresses or potential for microcrack propagation. Similarly, temporal gradients in surface temperature have been widely used as a proxy for efficacy of thermal damage. These gradients are easy to measure in the field or laboratory, so the prevalence of their use as a proxy over the last few decades is understandable. It has been assumed that rapid changes in temperature at the grain (or thermocouple) scale can set up spatial temperature gradients that induce high stresses. However, our results show that these temporal gradients do not correlate with grain- or domain-scale stresses in the near surface. As we have shown, these stresses are not controlled by local temperature gradients, but rather by heterogeneity in elastic properties. We suggest that grain scale processes of a material can be better understood by analyzing the thermoelastic behavior of its mineral constituents. If a proxy is needed (e.g. in planetary work) to compare relative efficacy of thermomechanical processes, we suggest that the temperature or offset from the diurnal temperature mean be used, as they have direct control over induced stresses and we understand better how they will vary with microstructure, material, and environment.

## 5. Conclusions:

We believe these results provide very strong evidence for the significance of thermomechanical processes on airless bodies. While the stresses presented represent a lower limit, we believe they can inform our understanding of thermomechanical breakdown on airless bodies. These methods are useful in constraining relative efficacy between bodies, but future work is needed to determine material strengths and actual availability of stresses for crack propagation in these environments, including understanding the relationship between grain and macroscopic scale processes in complex landscapes. Most importantly, this works shows:

1) The peak stress experienced by a microstructure with any heterogeneity is at least comparable to the baseline stress value of the material with the highest coefficient of thermal expansion. Analyzing the thermoelastic behavior of individual mineral constituents provides the simplest lower limit approximation of expected induced stress.

2) Average stresses in a microstructure are controlled by material properties and relative volume of mineral types, but the formation and propagation of cracks through that microstructure is controlled by grain and pore distribution.

3) Tips of cracks oriented perpendicular to the surface experience higher stresses, making vertical propagation of cracks energetically favorable. However, propagation will also be affected by local stress fields from the presence and distribution of mineral heterogeneity.

4) Grain scale stresses are controlled by the difference in thermal expansion coefficient between mineral grains, and to a lesser extent by Young's modulus and Poisson's ratio. Thermal conductivity, and thus the spatial temperature gradient, does not play a role at this scale.

5) Spatiotemporal grain or multi-grain scale temperature gradients are not an appropriate proxy for stresses induced at those scales. Temperature or offset from the diurnal temperature mean are more useful.

As we have shown in this study, analyzing the thermoelastic behavior of simple microstructures and mineral constituents is an informative approximation. Future work is needed to understand the more complex aspects of thermoelastic behavior of planetary materials. Additional modeling and laboratory work (especially in vacuum) investigating material strengths and behaviors are needed to quantitatively approximate breakdown rates in a more realistic way.

**Acknowledgments**

Support for this work came from the NASA Earth and Space Science Fellowship and the Planetary Geology and Geophysics programs. Additionally, this work would not have been possible without free access to OOF2 from the National Institute of Standards and Technology. We thank the OOF developers, as well as those who reviewed this manuscript. The data for this paper are available upon request by emailing the corresponding author (jmolaro@gmail.com).

----------------------------------------------------

**APPENDIX**

**A1. Material Properties**

Table A1.

| DENSITY | | | | |
|---|---|---|---|---|
| **Pyroxene** | | $\rho$ (kg/m$^3$) | | |
| | Enstatite | 3100 | Horai 1971 | |
| | Diopside | 3291 | Horai 1971 | |
| | Diopside | 3140 | Skinner 1966 | |
| | **VALUE USED** | **3275** | | |
| **Plagioclase** | | | | |
| | Plagioclase | 2630 | Skinner 1966 | |
| | Anorthite | 2699 | Schilling 2001 | |
| | **VALUE USED** | **2665** | | |

Table A2.

| YOUNG'S MODULUS | | | | |
|---|---|---|---|---|
| **Pyroxene** | | $E$ (GPa) | $\upsilon$ | |
| | Orthopyroxene | 183.55 | 0.2204 | Webb and Jackson 1993 |
| | Diopside | 167.07 | 0.2468 | Levien 1979 |
| | **VALUE USED** | **175** | **0.23** | |
| **Plagioclase** | | | | |
| | Anorthite | 107.43 | 0.3101 | Angel 1988, Lieberman and Ringwood 1976, Blundy and Wood 1990 |
| | Albite | 62.19 | 0.3417 | Angel 1988, Lieberman and Ringwood 1976, Blundy and Wood 1990 |
| | **VALUE USED** | **85** | **0.33** | |

Table A3.

| VOLUMETRIC COEFFICIENT OF THERMAL EXPANSION | | | |
|---|---|---|---|
| **Pyroxene** | | α (x$10^{-5}$) | |
| | Enstatite | 3.1 | Saxena 1988 |
| | Augite | 1.8 | Skinner 1966 |
| | Diopside | 2.4 | Skinner 1966 |
| | **VALUE USED** | **2.4** | |
| **Plagioclase** | | | |
| | AbAn(1) | 1.2 | Dante 1942, Skinner 1966 |
| | AbAn(2) | 1.2 | Dante 1942, Skinner 1966 |
| | AbAn(3) | 1.3 | Dante 1942, Skinner 1966 |
| | AbAn(4) | 1.8 | Dante 1942, Skinner 1966 |
| | **VALUE USED** | **1.2** | |

Table A4.

| THERMAL CONDUCTIVITY | | | |
|---|---|---|---|
| **Pyroxene** | | $k$ (W/m K) | |
| | Diopside* | 4.6 | Horai 1971 |
| | **VALUE USED** | **4.6** | |
| **Plagioclase** | | | |
| | Albite | 2.08 | Horai 1971 |
| | Anorthite | 1.88 | Horai 1971 |
| | **VALUE USED** | **2** | |

*While only one value of the thermal conductivity of pyroxene is referenced here, we found several references for the thermal conductivity of olivine. We found that the material property values for olivine were very similar to those of pyroxene, thus we determined the above represented an acceptable approximation of this parameter value.

Table A5.

| HEAT CAPACITY | | | |
|---|---|---|---|
| **Pyroxene** | | $c_P$ (J/kg K) | |
| | Diopside | 970 | Richet 1991 |
| | Orthopyroxene | 676 | Ashida 1988 |
| | **VALUE USED** | **823** | |
| **Plagioclase** | | | |
| | Albite | 1200 | Martens 1987 |
| | Anorthite | 962 | Richet 1991 |
| | **VALUE USED** | **1081** | |

**A2. Boundary Conditions**

This model represents a microstructure within a larger structure that can be represented as an infinite halfspace. The use of periodic boundary conditions for the displacement equation in the x-direction significantly increases the computational time needed to process results, especially in the more complex heterogeneous microstructures. A test was performed using a simpler microstructure (Figure A1a) with a domain size of 5 mm and a grain size of 0.9 mm (see Appendix Section A3). Profiles of the average and maximum stresses in this microstructure are shown in Figure A1b. The test with rigid boundary conditions (black) produced lower average stresses, and higher maximum stresses than the test with periodic boundary conditions (green). However, both models predicted average and maximum effective stresses during the state of peak tension to within ~1% of each other. We opted to use the faster, rigid boundary conditions in order to be able to simulate more cases.

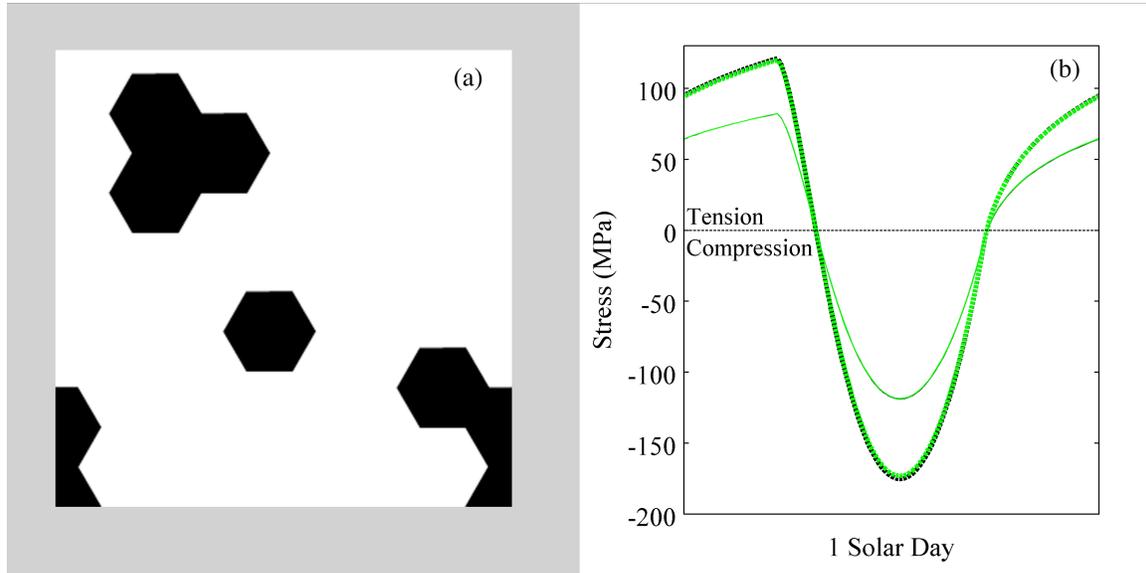

**Figure A1.** (a) The simple microstructure used to perform the boundary conditions tests. (b) Profiles of the average (solid) and maximum (dotted) stresses over time for model runs with rigid (black) and periodic (green) displacement boundary conditions. The two black lines are plotted, but are somewhat hidden beneath the green lines.

**A3. Effect of material properties on microstructures (I) and (VI)**
In the results presented, we found that the maximum stress induced in a homogeneous pyroxene microstructure (Table 2, I) is approximately equal to that induced in a plagioclase microstructure which contains a single pyroxene grain (Table 2, VI). This suggests that the lower limit stress any heterogeneous microstructure will experience is approximately equal to that of whichever component has higher elasticity and thermal expansion. To support this suggestion, we simulated cases using the same microstructures (Table 2, I and VI), but with differing Young's modulus, Poisson's ratio, and coefficient of thermal expansion (Table A6). Since the difference in elastic properties controls the magnitude of stresses, we started with those values as originally defined in the main study, and then increased how different each parameter is between the two mineral types. For example, the original values of the Young's modulus for pyroxene and plagioclase were 175 and 85 MPa, respectively, giving a difference of 90 MPa. We increased the value for pyroxene to 220 MPa, and decreased the value for plagioclase to 40 MPa, giving a difference of 180 MPa. This allows us to determine how comparable stresses induced in microstructures I and VI are for multiple sets of elastic properties.

This test tells us is how close the stress on a homogeneous microstructure ($\sigma_I$) is to the lower limit stress expected in a heterogeneous microstructure ($\sigma_{VI}$). The results of these tests are shown in Table A7. The first two rows show the maximum stress on microstructures of type I and VI ($\sigma_I$ and $\sigma_{VI}$, respectively). The third row is the difference between those two values ($\Delta\sigma$), and the last row is the percentage of $\sigma_I$ that difference represents. This percentage is a measure of how accurate that lower limit approximation is. The results indicate that in all cases tested, $\sigma_I$ varies from ~8-18% higher than $\sigma_{VI}$. The actual percentage expected depends on the mineral properties of whatever microstructure is being simulated or approximated. However, since the range tested here goes beyond typical values for pyroxene and plagioclase, we suggest using $\sigma_I$ as an approximation of $\sigma_{VI}$ for these microstructures is accurate within ~18%. Microstructures containing minerals with significantly different properties than pyroxene and plagioclase may require additional testing.

No appreciable changes in stress were detected in tests conducted on microstructures (type I) with varying grain sizes. Thus while no specific tests were conducted with the new properties in Table A6, we believe the role of grain size (or domain size) on this approximation to be negligible.

Table A6.

| Material Property | Pyroxene | Plagioclase |
|---|---|---|
| Original Case | | |
| $E$ (GPa) | 175 | 85 |
| $\nu$ | 0.23 | 0.33 |
| $\alpha$ (K$^{-1}$) | 0.8x10$^{-5}$ | 0.4x10$^{-5}$ |
| Test 1 | | |
| $E$ (GPa) | 190 | 70 |
| $\nu$ | 0.2133 | 0.3467 |
| $\alpha$ (K$^{-1}$) | 0.867x10$^{-5}$ | 0.333x10$^{-5}$ |
| Test 2 | | |
| $E$ (GPa) | 205 | 55 |
| $\nu$ | 0.1966 | 0.3634 |
| $\alpha$ (K$^{-1}$) | 0.934x10$^{-5}$ | 0.266x10$^{-5}$ |
| Test 3 | | |
| $E$ (GPa) | 220 | 40 |
| $\nu$ | 0.18 | 0.38 |
| $\alpha$ (K$^{-1}$) | 1x10$^{-5}$ | 0.2x10$^{-5}$ |

Table A7.

| Microstructure | Original | Test 1 | Test 2 | Test 3 |
|---|---|---|---|---|
| Peak Stress: $\sigma_I$ (MPa) | 97 | 109 | 124 | 139 |
| Peak Stress: $\sigma_{VI}$ (MPa) | 84 | 93 | 106 | 128 |
| $\Delta\sigma_{I-VI}$ (MPa) | 13 | 16 | 18 | 11 |
| % of $\sigma_I$ | 13.4 | 17.4 | 14.5 | 7.9 |

## A4. Effect of grain distribution on induced stresses

We simulated cases on multiple microstructures with the properties of the standard case (Table 2, III) to determine the effect of grain distribution on maximum stresses induced. Figure A2 shows a snapshot of three different microstructures during the state of peak tension. As in the results presented in the main text, amplification of stresses occurs where intergrain boundaries are clustered. Figure A3 shows a profile of the maximum stress for each microstructure in Figure A2 over time. While there is some variation in magnitude, maximum stresses in each case are within 17 MPa of each other. As such, we have determined that particular grain distribution will not affect our conclusions, especially given the fact that material strengths are not well defined.

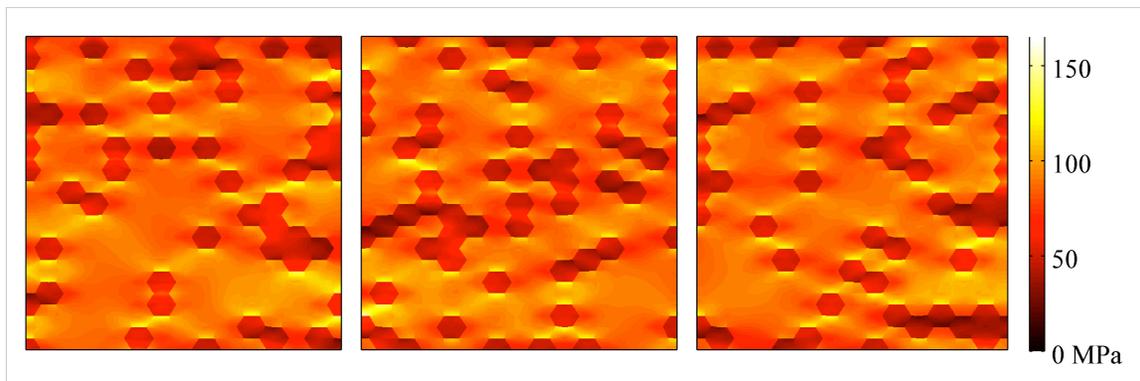

**Figure A2.** Snapshots of the state of peak tensile stress in three microstructures with the properties of the standard microstructure (Table 2 III). Panel (a) is the standard case used in the main text (Figure 5, center panel). Panels (b) and (c) have the same properties but their grains have different random distributions.

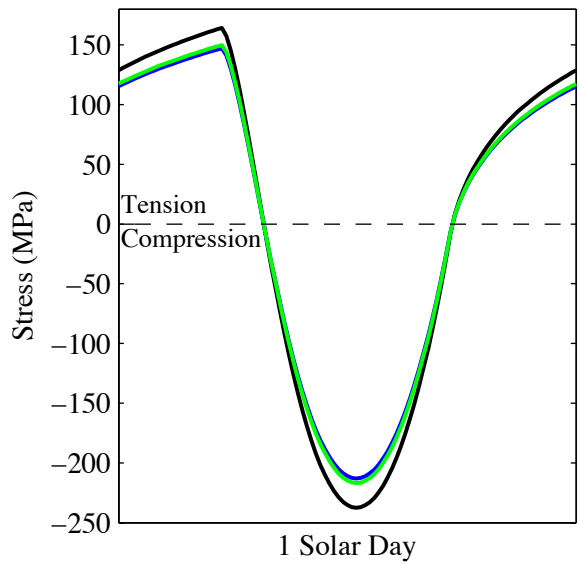

**Figure A3.** Profiles of maximum stresses induced in three microstructures (Figure A2) with equal relative mineral volume, but differing distribution. The green line corresponds to Figure A2a, black to A2b, and blue to A2c.